\documentclass[aps,showpacs,twocolumn,superscriptaddress]{revtex4-1}
\let\revappendix\appendix

\usepackage{graphicx} 
\usepackage{color}
\usepackage{array}
\usepackage{mathrsfs}

\usepackage{ulem} 

\begin{document}

\title{Magnetization plateaus in the spin-$\frac{1}{2}$ antiferromagnetic Heisenberg model on a kagome-strip chain}



\author{Katsuhiro Morita}
\email[e-mail:]{katsuhiro.morita@rs.tus.ac.jp}
\affiliation{Department of Applied Physics, Tokyo University of Science, Tokyo 125-8585, Japan}

\author{Takanori Sugimoto}
\affiliation{Department of Applied Physics, Tokyo University of Science, Tokyo 125-8585, Japan}

\author{Shigetoshi Sota}
\affiliation{RIKEN Advanced Institute for Computational Science (AICS), Kobe, Hyogo 650-0047, Japan}

\author{Takami Tohyama}
\affiliation{Department of Applied Physics, Tokyo University of Science, Tokyo 125-8585, Japan}


\date{\today}

\begin{abstract}
The spin-$\frac{1}{2}$ Heisenberg model on a kagome lattice is a typical frustrated quantum spin system. The basic structure of a kagome lattice is also present in the kagome-strip lattice in one dimension, where a similar type of frustration is expected.
We thus study the magnetization plateaus of the spin-$\frac{1}{2}$ Heisenberg model on a kagome-strip chain with three-independent antiferromagnetic exchange interactions using the density-matrix renormalization group method. In a certain range of exchange parameters, we find twelve kinds of magnetization plateaus, nine of which have magnetic structures breaking translational and/or reflection symmetry spontaneously. The structures are classified by an array of five-site unit cells with specific bond-spin correlations. In a case with a nontrivial plateau, namely a 3/10 plateau, we find long-period magnetic structure with a period of four unit cells.
\end{abstract}


\maketitle
\section{Introduction}
Quantum phase transitions are a subject undergoing intense study in the field of condensed-matter physics. In geometrically frustrated quantum spin systems, quantum phase transitions are frequently induced by applying a magnetic field. At zero temperature, magnetization plateaus, cusps, and jumps are manifestations of the transitions.

A typical frustrated system is a spin-$\frac{1}{2}$ two-dimensional (2D) Heisenberg model with a kagome lattice~\cite{Balents2010}. With the high magnetic field, the saturation of magnetization, $M_{\rm sat}$, occurs in this system. Upon decreasing the magnetic field, there is a sudden decrease in magnetization  $M$ from $M_{\rm sat}$ to a plateau with $M/M_{\rm sat}=7/9$, which is described by localized  multi-magnon states (LMMSs)~\cite{localmag1,localmag2}. Upon decreasing the magnetic field further, magnetization plateaus with $M/M_{\rm sat}=5/9$, 1/3, and 1/9 are predicted. However, no consensus has been reached on the magnetic structure of their ground state. For example, a valence-bond crystal (VBC) with $\sqrt{3} \times \sqrt{3}$ order~\cite{kagome1,kagome2,kagome-h1} and an up-up-down structure~\cite{kagome3} have been predicted for the ground state at the 1/3 plateau. The absence of the 1/3 plateau was also discussed~\cite{kagome-h2}. For the 1/9 plateau, the ground state has been proposed as either Z$_3$ spin liquid~\cite{kagome1} or a VBC~\cite{kagome3}. Even for zero magnetic field, the nature of the ground state is still under discussion: it is either a gapped Z$_2$ spin liquid~\cite{Z2-1,Z2-2}, a gapless U(1) spin liquid~\cite{U1-1,U1-2,U1-3,U1-4}, or a VBC~\cite{VBC-1,VBC-2,VBC-3}. Therefore, the understanding of magnetization processes and induced quantum phase transitions in the kagome-type 2D Heisenberg model is still far from complete.

\begin{figure}[tb]
\includegraphics[width=70mm]{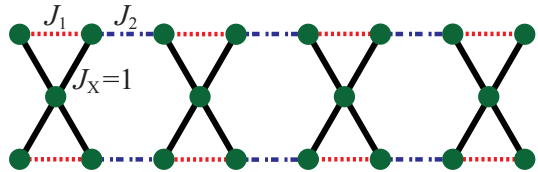}
\caption{Structure of a kagome-strip chain. The black solid, red dashed, and blue broken lines denote the exchange interactions $J_{\rm X}$, $J_1$, and $J_2$, respectively. We set $J_{\rm X}$=1.
\label{lattice}}
\end{figure}

Another kagome-type frustrated model is a spin-$\frac{1}{2}$ one-dimensional (1D) Heisenberg model with a kagome strip (see Fig.~\ref{lattice}). Since the model contains a basic kagome structure, i.e., a five-site unit cell, the clarifications of magnetization processes and field-induced quantum phase transition in the model may contribute to our further understanding of the magnetic properties of the 2D kagome lattice. The ground state of the spin-$\frac{1}{2}$ kagome-strip chain has been studied, and it was realized that a strongly localized Majorana fermion will exist in zero magnetic field~\cite{ksc1}. The singlet-triplet gap has been estimated to be around 0.01$J$ in the condition in which all of the exchange interactions are equivalent~\cite{ksc2}, i.e., $J_{\rm X}=J_1=J_2$ in Fig.~\ref{lattice}. However, the ground state of this model in the magnetic field has not been examined and thus detailed characteristics in the field are not known. Furthermore, the model containing inequivalent three exchange interactions has not been examined as far as we know even for zero magnetic field. 
Recently, a compound with a distorted kagome-strip chain, $A_2$Cu$_5$(TeO$_3$)(SO$_4$)$_3$(OH)$_4$ ($A$ = Na, K), was reported~\cite{kscex}. Therefore, the model is now attractive not only for a purely theoretical investigation but also for an experimental one.

In this paper, we study the ground state of a spin-$\frac{1}{2}$ Heisenberg model on the kagome-strip chain in a magnetic field using the density matrix renormalization group (DMRG) method. We accurately determine the magnetic structures of the chain in a certain parameter range, and we find various types of plateaus that have not been reported before. In total, we identify twelve kinds of magnetization plateaus, nine of which have magnetic structures that break translational and/or reflection symmetry spontaneously. The structures are classified by an array of five-site unit cells with specific bond-spin correlations. Among the plateaus, we find a nontrivial plateau, namely a 3/10 plateau, whose magnetic structure consists of a period of four unit cells. To the best of our knowledge, such long-period magnetic structure has not been reported before in 1D quantum spin systems.

The arrangement of this paper is as follows.  In Sec.~\ref{sec:2}, we describe our kagome-strip chain model and numerical method. In Sec.~\ref{sec:3}, we discuss the results of the magnetization plateaus and magnetic structures.  Finally, a summary is given in Sec.~\ref{sec:4}.

\section{model and method}
\label{sec:2}

\begin{figure}[tb]
\includegraphics[width=86mm]{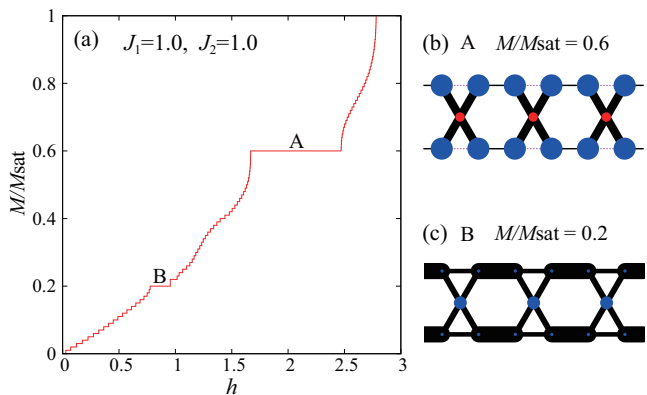}
\caption{(a) Magnetization curve of the kagome-strip chain with $N=200(=5\times40)$ for $J_{\rm X}$=$J_{\rm 1}$=$J_{\rm 2}$=1 at zero temperature. The symbol A (B) represents the 3/5 (1/5) plateau.
(b) The nearest-neighbor spin correlation $\langle \mathbf{S}_i\cdot\mathbf{S}_j\rangle$ -$\langle S^z_i\rangle$$\langle S^z_j\rangle$ and the local magnetization $\langle S^z_i\rangle$  in the 3/5 plateau for  $5\times3$ sites at the center of the chain. 
Black solid (purple dashed) lines connecting two nearest-neighbor sites denote negative (positive) values of the spin correlation, and their thicknesses represent the magnitudes of correlation. 
Blue (red) circles on each site denote a positive (negative) value of $\langle S^z_i\rangle$, and their diameters represents its magnitude.
(c) Same as (b) but for the 1/5 plateau. 
\label{m-h-1}}
\end{figure}

The Hamiltonian for a spin-$\frac{1}{2}$ kagome-strip chain in a magnetic field is defined as
\begin{eqnarray} 
H &=& \sum_{\langle i,j \rangle }J_{i,j} \mathbf{S}_i \cdot \mathbf{S}_j - h\sum_i S^{z}_i,
\label{Hami}
\end{eqnarray}
where $\mathbf{S}_i$ is the spin-$\frac{1}{2}$ operator, $\langle i,j \rangle$ runs over the nearest-neighbor spin pairs, $J_{i,j}$ corresponds to one of $J_{\rm X}$, $J_1$, and $J_2$ shown in Fig.~\ref{lattice}, and $h$ is the magnetic field magnitude. In the following, we set $J_{\rm X}=1$ as an energy unit. We perform the DMRG calculations at zero temperature for the kagome-strip chain up to system size $N=325(=5\times65)$ in the open boundary condition (OBC) for various values of $J_1$ and $J_2$. The number of states kept in the DMRG calculations are $m =400$, and resulting truncation errors are less than $5 \times 10^{-7}$.
Since the chain is formed by an array of five-site unit cells, $N$ is a multiple of 5. Furthermore, we determine the number of unit cells by taking into account a period of magnetic structures in each plateau.

\section{results and discussion}
\label{sec:3}
We first consider equivalent exchange interactions, i.e., $J_{\rm X}=J_1=J_2=1$, for comparison with the 2D kagome lattice. Figure~\ref{m-h-1}(a) shows the magnetization curve, where plateaus with  $M/M_{\rm sat}=$3/5 and 1/5 are observed. In addition, there is a shoulder-like anomaly around $M/M_{\rm sat}=$2/5, which looks like the signature of the 2/5 plateau. We do not find a 4/5 plateau corresponding to the 7/9 plateau in the 2D Kagome lattice, indicating no LMMSs in the kagome-strip chain with equivalent exchange interactions. We cannot confirm a zero-magnetization plateau in our calculation because the singlet-triplet gap reported to be small (around 0.01$J$)~\cite{ksc2} may become even smaller due to the influence of OBC.

\begin{figure}[tb]
\includegraphics[width=64mm]{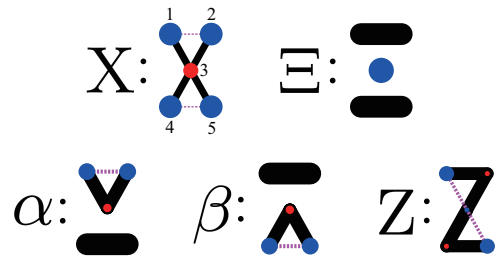}
\caption{Magnetic structures of a five-site $J_{\rm X}$-$J_1$ unit. Black solid (purple dashed) lines connecting two nearest-neighbor sites denote negative (positive) values of spin correlation defined as $\langle \mathbf{S}_i\cdot\mathbf{S}_j\rangle$ -$\langle S^z_i\rangle$$\langle S^z_j\rangle$, and their thicknesses represent the magnitudes of correlation. Blue (red) circles on each site denote a positive (negative) value of local magnetization $\langle S^z_i\rangle$, and their diameters represent its magnitude. We use symbols ``X'', ``$\Xi$'', ``$\alpha$'', ``$\beta$'', and ``Z'' to represent each magnetic structure. The total spin $S$ and its $z$-component $S_z$ in each state are equal, i.e., $S=S_z$, with 3/2, 1/2, 1/2, 1/2, and 1/2 for ``X'', ``$\Xi$'', ``$\alpha$'', ``$\beta$'', and ``Z'' respectively. The state ``Z'' is given by a linear combination of ``$\Xi$'', ``$\alpha$'', and ``$\beta$'' (see the Appendix).
\label{5site}}
\end{figure}

Figures~\ref{m-h-1}(b) and ~\ref{m-h-1}(c) show nearest-neighbor spin correlation $\langle \mathbf{S}_i\cdot\mathbf{S}_j\rangle$ -$\langle S^z_i\rangle$$\langle S^z_j\rangle$ and local magnetization $\langle S^z_i\rangle$ in the 3/5 and 1/5 plateaus, respectively. The lines connecting two nearest-neighbor sites denote the sign and magnitude of spin correlation by color and thickness, respectively. The circle on each site represents $\langle S^z_i\rangle$. We find that both spin correlation and magnetization shown in Figs.~\ref{m-h-1}(b) and \ref{m-h-1}(c) hold translational and reflection symmetries. This is in contrast to the 1/3, 5/9, and 7/9 plateaus of the 2D kagome lattice, breaking these symmetries~\cite{kagome1,kagome2,kagome3}.

\begin{figure}[tb]
\includegraphics[width=86mm]{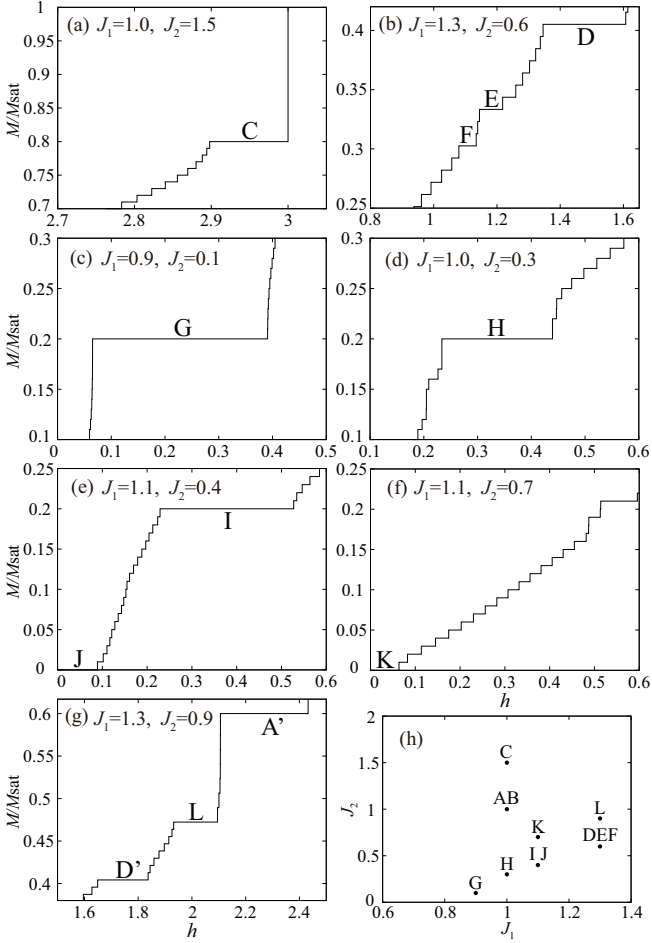}
\caption{Magnetization curve of the kagome-strip chain with OBC for $J_{\rm X}=1$ at zero temperature. (a) $J_1=1.0$ and $J_2=1.5$ for $N=200(=5\times40)$. (b) $J_1=1.3$ and $J_2=0.6$ for $N=195(=5\times39)$. (c) $J_1=0.9$ and $J_2=0.1$ for $N=200$. (d) $J_1=1.0$ and $J_2=0.3$ for $N=200$. (e) $J_1=1.1$ and $J_2=0.4$ for $N=200$. (f) $J_1=1.1$ and $J_2=0.7$ for $N=200$. (g) $J_1=1.3$ and $J_2=0.9$ for $N=235(=5\times 47)$. Magnetic structures of the plateaus, C, D, E, F, G, H, I, J, K and L shown in each panel,  are shown in Fig.~\ref{structure}. Magnetic structures of the plateaus, A' and D', are almost same as that of the plateau A in Fig.~\ref{m-h-1}(a) and D, respectively. (h) The position of the plateaus in the $J_2$ v.s. $J_1$ plane.
\label{m-h}}
\end{figure}

The magnetic structure in the 3/5 plateau consists of five-site clusters with ``X''-like structure as shown in Fig.~\ref{m-h-1}(b).  The ``X''-like magnetic structure is obtained in an eigenstate of a five-site unit with $J_{\rm X}$ and $J_1$ in the space of total spin $S=3/2$ and its $z$ component $S_z=3/2$ (see Appendix). The corresponding magnetic structure is shown in Fig.~\ref{5site}. The ``X'' state has spin correlation and local magnetization similar to a five-site unit cell in Fig.~\ref{m-h-1}(b).

We can construct other magnetic structures from eigenstates of the five-site $J_{\rm X}$-$J_1$ unit as denoted by ``$\Xi$'', ``$\alpha$'', ``$\beta$'', and ``Z'' for the space of $S=S_z=1/2$. These components appear in magnetic structures of magnetization plateaus in the kagome-strip chain as discussed below. Note that there is no five-site component corresponding to the magnetic structure in Fig.~\ref{m-h-1}(c), where the $J_2$ bond has the strongest spin correlation resembling the ground state in the limit $J_2 \rightarrow\infty$.

\begin{figure}[tb]
\includegraphics[width=86mm]{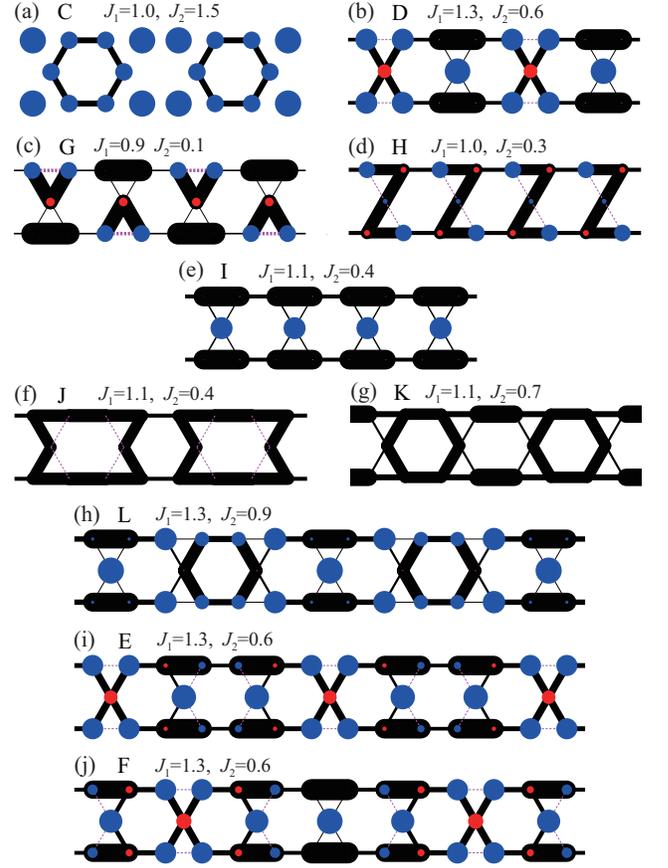}
\caption{The nearest-neighbor spin correlation $\langle \mathbf{S}_i\cdot\mathbf{S}_j\rangle$ -$\langle S^z_i\rangle$$\langle S^z_j\rangle$ and the local magnetization $\langle S^z_i\rangle$  in each plateau for  $5\times7$ sites at the center of the chain. (a) The plateau C, (b) D, (c) G, (d) H, (e) I, (f) J, (g) K, (h) L, (i) E, and (j) F in Fig.~\ref{m-h}. Black solid (purple dashed) lines connecting two nearest-neighbor sites denote negative (positive) values of the spin correlation and their thicknesses represents the magnitudes of correlation. Blue (red) circles on each site denote positive (negative) value of $\langle S^z_i\rangle$ and their diameters represents its magnitude.
\label{structure}}
\end{figure} 

We examine magnetization processes for various parameter sets of exchange interactions in $J_1$ vs. $J_2$ space as denoted by dots in Fig.~\ref{m-h}(h). At the equivalent case, i.e., $J_1=J_2=1$, we do not find the 4/5 plateau, but with increasing $J_2$ we identify the 4/5 plateau denoted by C in Fig.~\ref{m-h}(a) with $J_1=1$ and $J_2=1.5$. Not only the 4/5 plateau but also a macroscopic magnetization jump just below the saturation field is observed. The 4/5 plateau exhibits the wave function with a period of $5\times2$ as shown in Fig.~\ref{structure}(a), which indicates LMMS with a spontaneous translational symmetry breaking as expected from the 2D kagome lattice~\cite{localmag1}.

\begin{figure}[tb]
\includegraphics[width=70mm]{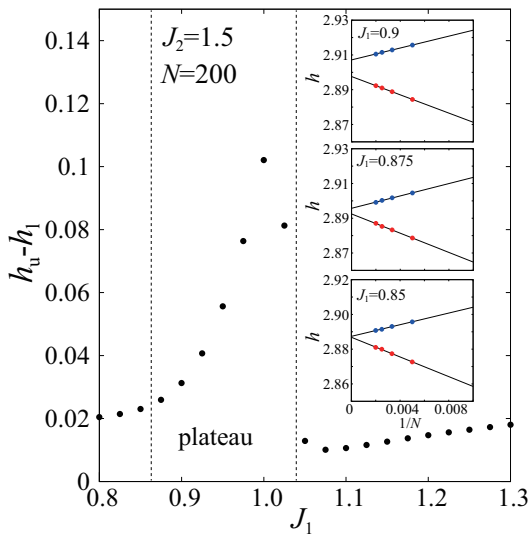}
\caption{The difference of the upper and lower magnetic fields, $h_\mathrm{u}-h_\mathrm{l}$, for $M/M_\mathrm{sat}=4/5$ as a function of $J_1$ at $J_2=1.5$ for $N=200$. The 4/5 plateau exists at $0.86\lesssim J_1 \lesssim 1.03$. Inset: the size scaling of $h_\mathrm{u}$ and $h_\mathrm{l}$ for several value of $J_1$ as a function of $1/N$. Both $h_\mathrm{u}$ and $h_\mathrm{l}$, denoted by blue and red circles, respectively, are extrapolated by a linear line.
\label{p-w}}
\end{figure}

We also examine the parameter region of $J_1$ at $J_2=1.5$, where the 4/5 plateau emerges. We calculate the lower magnetic field $h_\mathrm{l}$ and the upper magnetic field $h_\mathrm{u}$ for $M/M_\mathrm{sat}=4/5$. The difference $h_\mathrm{u}-h_\mathrm{l}$ is shown in Fig.~\ref{p-w} as a function of $J_1$ for $N=200$.  Because of the finite-size effect, $h_\mathrm{u}-h_\mathrm{l}$ is alway finite in spite of the absence of the plateau. At $J_1=1.03$, there is a jump of $h_\mathrm{u}-h_\mathrm{l}$, indicating the upper bound of the 4/5 plateau. On the other hand, there is no clear jump for the lower bound. To determine the boundary, we make a finite-size scaling for both $h_\mathrm{u}$ and $h_\mathrm{l}$, which is shown in the inset of  Fig.~\ref{p-w}. From the scaling, we estimate the lower boundary to be around $J_1=0.86$.

The 2/5 plateau is also missing in the equivalent case. With $J_1=1.3$ and $J_2=0.6$, however, we find the 2/5 plateau as denoted by D in Fig.~\ref{m-h}(b). The magnetic structure in this plateau is shown in Fig.~\ref{structure}(b), where an alternating order of five-site ``X''-like and ``$\Xi$''-like structures emerges. This phase also breaks translational symmetry.
We can understand the emergence of the 2/5 plateau for $J_1>1$ by using a first-order perturbation with respect to $J_2$. Based on the fact that, for $J_1>1$ the lowest energy state of the five-site cluster with $S_z=1/2$ is the ``$\Xi$'' state and that with $S_z=3/2$ is the ``X'' state, we can construct an effective Hamiltonian in the first order of $J_2$ given by
\begin{eqnarray} 
H_\mathrm{eff} &=& \frac{81}{200}J_2\sum_{\langle i,j \rangle }{T}_i^z{T}_j^z  \nonumber \\
&+&  \left(\frac{81}{200}J_2  +2J_1 - h -\frac{3}{2}\right)\sum_{i}{T}_i^z + \mathrm{const},
\label{eff1}
\end{eqnarray}
where ${T}_i^z$ represents the z component of the spin-$\frac{1}{2}$ operator at site $i$ acting on the ``$\Xi$'' and  ``X'' states. The effective Hamiltonian (\ref{eff1}) corresponds to an Ising single chain in a magnetic field. The N\'eel state is realized in the chain when $2J_1 -3/2 < h < 2J_1 -3/2 + \frac{81}{100}J_2$, which is equivalent to the alternating order of the ``X''-like and ``$\Xi$''-like structures in the 2/5 plateau obtained from the Hamiltonian (\ref{Hami}).

As shown in Fig.~\ref{m-h-1}(a), the 1/5 plateau denoted by B exists in the equivalent case $J_1=J_2=1$. In addition to B, we find three new types of the 1/5 plateau for  $J_1\sim1$ and small $J_2$. Plateaus in Figs.~\ref{m-h}(c), \ref{m-h}(d), and \ref{m-h}(e) denoted by G, H, and I, respectively, correspond to the new types. The magnetic structures for G, H, and I are shown in Figs.~\ref{structure}(c), \ref{structure}(d), and \ref{structure}(e), respectively. They are different from the magnetic structure in B as shown in Fig.~\ref{m-h-1}(a), in the sense that the spin correlation connecting two five-site units is so small that the component of magnetic structure is composed of five-site units with four types ``$\alpha$'', ``$\beta$'', ``$\Xi$'', and ``Z'' shown in Fig.~\ref{5site}. We note that the ``$\alpha$'', ``$\beta$'', and ``$\Xi$'' states are eigenstates of the five-site system, while the ``Z'' state is a linear combination of the three-state system (see the Appendix). At $0.5<J_1<1$, the ground state of the $S=S_z=1/2$ space in the five-site unit has two-fold degeneracy given by the ``$\alpha$'' and ``$\beta$' states. Introducing small $J_2$ [$J_2=0.1$ in Fig.~\ref{m-h}(c)], the degeneracy is lifted and, as a result, the plateau G exhibits alternating order of the ``$\alpha$'' and ``$\beta$'' states as shown in Fig~\ref{structure}(c). 
We can easily obtain an Ising-type effective Hamiltonian by the first-order perturbation with respect to $J_2$, which is given by
\begin{equation} 
H_\mathrm{eff} = \frac{2}{9}J_2\sum_{\langle i,j \rangle }\tilde{S}_i^z\tilde{S}_j^z + \mathrm{const},
\end{equation}
where $\tilde{S}_i^z$ represents the z component of the spin-$\frac{1}{2}$ operator at site $i$ acting on the ``$\alpha$'' and ``$\beta$'' state. When $J_2>0$, the ground states become the alternating  the ``$\alpha$'' and ``$\beta$'' order states, being consistent with the plateau G. Note that when $J_2<0$, the ground state is ferromagnetic corresponding either to the ``$\alpha$'' or ``$\beta$'' order states.
At $J_1=1$, the ``$\Xi$'' state degenerates with ``$\alpha$'' and ``$\beta$'' in the five-site unit. With small $J_2$  [$J_2=0.3$ in Fig.~\ref{m-h}(d)],  the degeneracy is lifted and the linear combination of these three states emerges in the plateau H, whose magnetic structure is an array of ``Z'' structures as shown in Fig.~\ref{structure}(d). At $J_1>$1, the ${\rm \Xi}$ state in Fig.~\ref{5site} becomes the ground state of the five-site unit and its array appears in the plateau I [see Figs.~\ref{m-h}(e) and \ref{structure}(e)].

We find two kinds of zero magnetization plateaus, the plateau J in Fig.~\ref{m-h}(e) for $J_1=1.1$ and $J_2=0.4$ and the plateau K in Fig.~\ref{m-h}(f) for $J_1=1.1$ and $J_2=0.7$, whose ground states are VBC with two-unit cells as shown in Figs.~\ref{structure}(f) and \ref{structure}(g), respectively. We find that magnetic structure forms decamers  [Fig~\ref{structure}(d)] or dimers and hexamers [Fig~\ref{structure}(e)]. The dimer-hexamer-ordered state is consistent with a previous report~\cite{ksc2}. However, there has been no report on the decamer-ordered state as far as we know. The presence of a large unit cell such as a decamer is surprising in the sense that VBC usually becomes unstable as the cluster size increases. This anomaly indicates strong frustration in this kagome-strip chain.

\begin{figure}[tb]
\includegraphics[width=86mm]{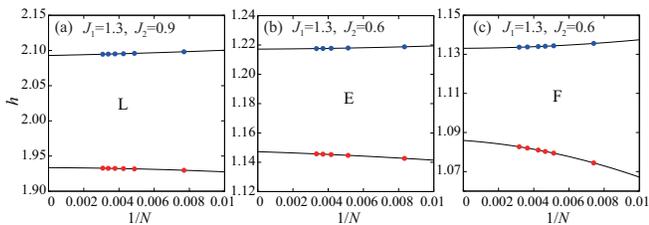}
\caption{Finite-size scaling of the lower and upper magnetic fields of magnetization plateaus. (a) The 7/15 plateau (L) in Fig~\ref{m-h}(g). (b) The 1/3 plateau (E) and (c) the 3/10 plateau (F) in Fig~\ref{m-h}(e). 
The solid lines show fitted results with scaling functions of $h_{\rm u(l)}= a_{\rm u(l)}(1/N)^2+b_{\rm u(l)}(1/N)+c_{\rm u(l)}$ for the upper (lower) field.
\label{1/N}}
\end{figure}

In the 1D systems consisting of only nearest-neighbor interactions, the appearance of long-period magnetic structures is a nontrivial phenomenon. We find such structures in the 7/15 (L) plateau in Fig.~\ref{m-h}(g) as well as the 1/3 (E) and 3/10 (F) plateaus in Fig.~\ref{m-h}(b). The magnetic structures of L, E, and F have periods of three-,  three-, and four-unit cells as shown in Figs.~\ref{structure}(h), \ref{structure}(i), and \ref{structure}(j), respectively. Magnetic structure of L plateau consists of almost upward spins, dimers and hexamers, while the structures of E and F are based on ``X'' and ``$\Xi$''. Since such long-period structures are usually unstable in the 1D systems, we perform finite-size scaling on the lower and upper magnetic fields of the three plateaus as shown in Fig.~\ref{1/N}. We set the scaling function of the upper (lower) fields $h_{\rm u(l)} = a_{\rm u(l)}(1/N)^2 + b_{\rm u(l)}(1/N) + c_{\rm u(l)}$. In all plateaus, L, E, and F, the difference of the upper and lower fields, $c_{\rm u}-c_{\rm l}$, is finite in the thermodynamic limit $N\rightarrow\infty$. This is an evidence of stable long-period structures in the three plateaus. 

We finally discuss the relationship between the magnetization plateaus we have found and the Oshikawa-Yamanaka-Affleck(OYA) criterion~\cite{OYA}.
In the kagome strip chain, the OYA criterion states that a necessary condition for the appearance of a plateau is $\frac{5}{2}p(1-M/M_{\rm sat}) = n$ ($n$: integer), where $p$ is the ground state period based on the unit cell. All plateaus obtained in the present study satisfy the OYA criterion. For example,  the $M/M_{\rm sat}=3/10$ plateau (F) is compatible with $p=4$, and the $M/M_{\rm sat}=0$ plateaus (J and K) are compatible with $p=2$.

\section{summary}
\label{sec:4}
In summary, motivated by recent progress in our understanding of frustration in the 2D kagome lattice, we have investigated the ground state of a spin-$\frac{1}{2}$ Heisenberg model on the kagome-strip chain in magnetic field using the DMRG method. We have accurately determined the magnetic structures of the chain in a certain parameter range, and we have found various types of plateaus that have not been reported before. We have identified twelve kinds of magnetization plateaus, nine of which have magnetic structures that break translational and/or reflection symmetry spontaneously. Among the nine plateaus, we have identified a nontrivial plateau, a 3/10 plateau, whose magnetic structure consists of a period of four unit cells. To the best of our knowledge, this is the first report of such a long-period magnetic structure in 1D quantum spin systems. 
All plateaus obtained in the present study satisfy the OYA criterion.
Our study reveals that there are a number of magnetization plateaus even with three different exchange interactions in the kagome-strip chain. It thus suggests that magnetization plateaus appear not only in a perfect kagome lattice~\cite{perkago} but even in distorted kagome lattices~\cite{dskago1,dskago2,dskago3,dskago4,dskago5,dskago6,dskago7,dskago8,dskago9,dskago10,dskago11}. 
In fact, a 1/3 plateau confirmed in Cs$_2$Cu$_3$CeF$_{12}$~\cite{dskago11} corresponds to the 1/5 plateaus, B and I, in the kagome-strip chain.
The recently discovered compound $A_2$Cu$_5$(TeO$_3$)(SO$_4$)$_3$(OH)$_4$ ($A$ = Na, K)~\cite{kscex} consists of a spin-$\frac{1}{2}$ kagome-strip chain. Though the compound has a distorted five-site unit cell in contrast to an undistorted cell in the present theoretical model, the kagome-strip chain, it might be a possible candidate to reveal nontrivial magnetization plateaus. Experimental and theoretical studies to confirm the plateaus in this compound are desired.

\begin{acknowledgments}
This work was supported in part by MEXT as a social and scientific priority issue [creation of new functional devices and high-performance materials to support next-generation industries (CDMSI) to be tackled by using a post-K computer and by MEXT HPCI Strategic Programs for Innovative Research (SPIRE) (hp160222)]. The numerical calculation was partly carried out at the K Computer, Institute for Solid State Physics, The University of Tokyo, and the Information Technology Center, The University of Tokyo. This work is also supported by a Grants-in-Aid for Scientific Research (No. 26287079) and a Grants-in-Aids for Young Scientists (B) (No.~16K17753) from MEXT, Japan.
\end{acknowledgments}
\revappendix*
\section{Five site model}
We discuss the eigenstates of a five-site $J_{\rm X}-J_1$ unit cell using an approach similar to that of Ref.~\cite{2d}. The Hamiltonian is defined as
\begin{eqnarray} 
H &=& J_{\rm X} \mathbf{S}_3 \cdot ( \mathbf{S}_1 + \mathbf{S}_2 + \mathbf{S}_4 + \mathbf{S}_5 )  + J_1 (\mathbf{S}_1 \cdot  \mathbf{S}_2 + \mathbf{S}_4 \cdot  \mathbf{S}_5)  \nonumber \\
	&=& J_{\rm X} \mathbf{S}_3 \cdot ( \mathbf{T}_u + \mathbf{T}_d ) + J_1 \left(  \frac{\mathbf{T_{\it u}^{\rm 2}}}{2} + \frac{\mathbf{T_{\it d}^{\rm 2}}}{2} \right)  - \frac{3}{2} J_1,
\end{eqnarray} 
where subscript numbers of $\mathbf{S}$ represent lattice position in Fig.~3 in the main text, and $\mathbf{T_{\it u}} = \mathbf{S}_1 + \mathbf{S}_2$ and $\mathbf{T_{\it d}} = \mathbf{S}_4 + \mathbf{S}_5$.
We can easily confirm that $[H,\mathbf{T}^2_{{\it u} {\rm (}{\it d}{\rm )}}]=0$ and $\mathbf{T}_{u {\rm (}d{\rm )}}^2=T_{u(d)}(T_{u(d)}+1)$ for $T_{u(d)}\in\{0,1\}$.
Each eigenstate of $H$ can be characterized by $T_u$ and $T_d$, and eigenstates with different set of $(T_u,T_d)$ are orthogonal to each other.

In the space of total spin $S=3/2$ and its z component $S_z=3/2$ for $(T_u,T_d)=(1,1)$, the ``X'' state shown in Fig.~3 in the main text appears. The eigenfunction of ``X'' is expressed as
\begin{eqnarray}
\nonumber |{\rm X}\rangle &=&  \frac{1}{\sqrt{10}} |t\rangle^+_{12}|\!\uparrow  \, \rangle_3|t\rangle^0_{45} -\frac{2}{\sqrt{5}}|t\rangle^+_{12}|\!\downarrow \,  \rangle_3|t\rangle^+_{45} \\
 &+& \frac{1}{\sqrt{10}}|t\rangle^0_{12}|\!\uparrow \,  \rangle_3|t\rangle^+_{45}, \label{X}
\end{eqnarray}
with eigenvalue $-\frac{3}{2}J_{\rm X}+\frac{1}{2}J_1$.
In the space of $S=S_z=1/2$ for  $(T_u,T_d)=(1,0)$, $(0,1)$, and $(0,0)$, the eigenfunctions of the ``$\alpha$'', ``$\beta$'', and ``$\Xi$'' states (see Fig.~3 in the main text) are given by 
\begin{eqnarray}
|{\alpha}\rangle &=&  \frac{1}{\sqrt{3}} \left[   \sqrt{2}|t\rangle_{12}^+  |\!\downarrow  \, \rangle_3   |s\rangle_{45} - |t\rangle_{12}^0 | \!\uparrow \,  \rangle_3 |s\rangle_{45}  \right],
\end{eqnarray}
\begin{eqnarray}
|{\beta}\rangle &=&  \frac{1}{\sqrt{3}} \left[ \sqrt{2} |s\rangle_{12}  |\!\downarrow  \, \rangle_3   |t\rangle_{45}^+ - |s\rangle_{12} | \!\uparrow \,  \rangle_3 |t\rangle_{45}^0 \right],
\end{eqnarray}
and
\begin{eqnarray}
|{\rm \Xi}\rangle =   |s\rangle_{12}|\!\uparrow  \, \rangle_3|s\rangle_{45}, \label{Xi}
\end{eqnarray}
with eigenvalues $-J_{\rm X}-\frac{1}{2}J_1$, $-J_{\rm X}-\frac{1}{2}J_1$, and $-\frac{3}{2}J_1$, respectively, where
\begin{eqnarray}
|t\rangle^+_{ij} &=&  |\!\uparrow \, \rangle_i|\!\uparrow \, \rangle_j , \\
|t\rangle^0_{ij} &=&  \frac{1}{\sqrt{2}}(|\!\uparrow \, \rangle_i|\!\downarrow \, \rangle_j  + |\!\downarrow \, \rangle_i|\!\uparrow \, \rangle_j) , \\
|s\rangle_{ij} &=&  \frac{1}{\sqrt{2}}(|\!\uparrow \, \rangle_i|\!\downarrow \, \rangle_j  - |\!\downarrow \, \rangle_i|\!\uparrow \, \rangle_j) .
\end{eqnarray}

At $J_{\rm X}=J_1$, the ``$\alpha$'', ``$\beta$'', and ``$\Xi$'' states are degenerate.  The ``Z''  state, which is a linear combination of the three states, is also the eigenstate of $H$ whose eigenfunction is given by
\begin{eqnarray}
|{\rm Z}\rangle &=& \frac{\sqrt{3}}{\sqrt{10}}|{\alpha}\rangle -\frac{\sqrt{3}}{\sqrt{10}}|{\beta}\rangle-\frac{2}{\sqrt{10}}|{\rm \Xi}\rangle \nonumber \\
&=&  \frac{2}{\sqrt{10}} \left[|\!\uparrow  \, \rangle_1  |s\rangle_{23}   |s\rangle_{45} + |s\rangle_{12}|s\rangle_{34} |\!\uparrow \,  \rangle_5 \right].
\end{eqnarray}

\end{document}